\documentclass[a4paper]{article}
\usepackage{setspace}
\usepackage{amsmath}
\usepackage{graphicx,subfigure}
\usepackage{epstopdf}
\usepackage{upgreek}
\usepackage[margin=2.0 cm]{geometry}

\begin{document}
\begin{center}

\large \textbf{Orientation Dependent Interaction and Self-assembly of Cubic Magnetic Colloids in a Nematic Liquid Crystal}

\hfill \break
Devika V S, Ravi Kumar Pujala, and Surajit Dhara$^{*}$

\hfill \break

\end{center}

\begin{abstract}
Spherical microparticles dispersed in nematic liquid crystals have been extensively investigated in the past years. Here, we report experimental studies on the elastic deformation, colloidal interaction and self-assembly of hematite microcubes with homeotropic surface anchoring in a nematic liquid crystal. We demonstrate that the colloidal interaction and self-assembly of cubic colloids are orientation dependent. In a notable departure from the conventional microspheres, the microcubes stabilise diverse structures, such as bent chains, branches, kinks and closed-loops. The microcubes reorient under rotating external magnetic field, thereby experiencing an elastic torque in the medium, which allows us to measure the magnetic moment through competition between elastic and magnetic torques. Our findings envisage that the faceted magnetic colloids in liquid crystals are potential for developing new functional magnetic materials with specific morphologies. \\\\\\\\\\\\\\\\
      
\end{abstract}

\
\noindent\rule{16cm}{0.5pt}\\
Devika V S and Prof. Surajit Dhara \\ 
  School of Physics, University of Hyderabad, Hyderabad-500046, India \\ 
 Dr. Ravi Kumar Pujala \\
Soft and Active Matter group, Department of Physics, Indian Institute of Science Education and Research, Tirupati-517507, India  \\
 $^{*}$Corresponding author:sdsp@uohyd.ernet.in \\

\newpage
\section{Introduction}
Self-assembly of colloidal particles into large ordered structures is a subject of immense interest in the field of soft matter, owing to the fundamental science and technological applications\textsuperscript{\cite{gm,noda,vw,av}}. In aqueous suspensions, the colloidal interaction is mostly driven by entropic or electrostatic forces, which are short-range and isotropic, thus limiting the diversity in the resulting colloidal structures. Colloidal dispersions in nematic liquid crystals (NLCs) are very promising as they induce variety of topological defects and interact via long-range elastic forces of the medium\textsuperscript{\cite{stark,poulin,sen,zuhail1,ivan,zumer,skarabot,zuhail2,zuhail3}}. They are especially interesting because the interaction is anisotropic and shape dependent, thus providing a route to targeted colloidal assemblies\textsuperscript{\cite{rasi,utk,laponite,rasna}}.
  Spherically symmetric particles with homeotropic surface anchoring nucleates either a point defect (hedgehog) or a disclination ring (Saturn ring)\textsuperscript{\cite{ramaswamy,stark}} in their vicinity. The colloids with accompanying point and ring defects are referred to as elastic dipoles and quadrupoles, based on the similarities of the director fields with their electric counterparts\textsuperscript{\cite{stark,igor}}. The defect mediated elastic interaction gives rise to two and three-dimensional colloidal structures and superstructures, bearing strong potential for photonic applications such as photonic bandgap crystals\textsuperscript{\cite{zumer,rav,ut,mar}}.  
  
\begin{figure}[ht]
\centering
\includegraphics[scale=0.25]{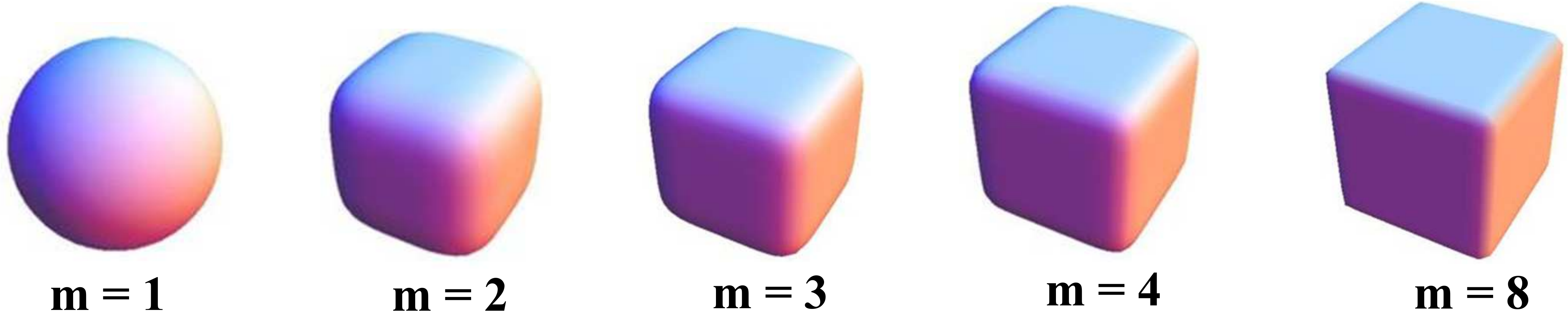}
\caption{Superellipsoids at various values of sharpness parameter $m$, showing the gradual change of roundness of the corners.} 
\label{fig:figure1}
\end{figure}

   Research on nematic colloids is mostly concentrated so far on spherical particles\textsuperscript{\cite{igor}}. Due to the advancement of synthesis and fabrication techniques, recently a surge has been observed in the number of investigations on nonspherical colloids, such as micro and nano rods\textsuperscript{\cite{utk,rasi}}, ellipsoids\textsuperscript{\cite{mta}}, peanuts\textsuperscript{\cite{dinesh}}, microbullets\textsuperscript{\cite{mag}}, circular and polygonal platelets\textsuperscript{\cite{laponite,rasna,jd}}, handle-bodies of varying genus~\textsuperscript{\cite{sen}} and Mobius strips\textsuperscript{\cite{tma}}. These investigations have provided ample of new results on the topological defects and ensuing self-assembled structures. However, experimental studies on the faceted colloids such as cubes with flat faces are largely unexplored. 
   There are of course a few simulation studies on microcubes in NLC showing the sharp edges act as pinning sites of Saturn ring defect\textsuperscript{\cite{beller,hung}}. The cube is modelled using a special case of the superellipsoid equation:  $(x)^{2m}+(y)^{2m}+(z)^{2m}=a^{2m}$, where $m$ is a shape parameter, defining the roundedness of the corners. It represents a sphere of radius $a$ for $m=1$, and a cube of side length $L=2a$ for $m\rightarrow\infty$ as shown in Fig.\ref{fig:figure1}. As $m$ interpolates the shape between a sphere and a cube, the defect loop evolves from a circular to distorted ring that wraps around the colloid while following only the edges\textsuperscript{\cite{beller}}.  
   
  In this paper, we for the first time report experimental studies on magnetic microcubes dispersed in a thin film of nematic liquid crystal. We study spontaneous orientation, elastic interaction and laser tweezers assisted colloidal assembly. We show that the cubic colloids stabilise diverse assemblies, which are not viable in spherical colloids. The cubic particles are made of hematite and hence respond to  external magnetic fields, thereby enabling us to determine the magnetic moment from the competing effects of magnetic and elastic torques. The magnetic response provides an additional degree of freedom for manipulation and controlled assembly of colloids in liquid crystals.
     
\section{Experiment}
 We synthesised cubic hematite particles ($\alpha$-Fe\textsubscript{2}O\textsubscript{3}) using sol-gel method\textsuperscript{\cite{sug1,sug2}}. In the synthesis, NaOH solution (100 ml, 6M) is slowly added to 2.0M  FeCl\textsubscript{3}  solution (100 ml) in a Pyrex bottle and agitated for 15 minutes. The entire solution is aged for 8 days at 100$^{\circ}$C and washed with milliQ water to obtain the hematite cubes. A thin layer (60 nm) of Silica (SiO\textsubscript{2}) coating on the cubes is made via an adaption of Stober method\textsuperscript{\cite{cg,jm}}.  Approximately 0.5g of hematite cubes is dispersed in 5 ml of water and resulting dispersion is mixed with 50 ml of polyvinylpyrrolidone (PVP) solution. The mixture is stirred for 8h and then sonicated to ensure the dissolution of PVP. Later, ethanol, water and TMAH solution is added, followed by the dropwise addition of  tetraethoxysilane (TEOS). The resulting solution is washed several times by centrifugation and redispersed in ethanol to obtain silica coated hematite microcubes. 
  Further, the silica coated microcubes are treated with N, N-dimetyl-N-octadecyl-3 aminopropyl-trimethoxysilyl chloride (DMOAP) to ensure perpendicular or homeotropic anchoring of the liquid crystal molecules on the surface. 
  
  In the next step, the microcubes are dispersed (1 wt.\%) in a room-temperature nematic liquid crystal, pentyl cyanobiphenyl (5CB) by using a vortex mixture and ultra sonicator. 
  The nematic colloid is filled in cells prepared by
  two parallel glass plates, which are polyimide (AL-1254) coated and baked at 180 $^{\circ}$C, followed by a unidirectional rubbing. The nematic director $\bf{n}$ (mean molecular orientation) is aligned along the rubbing direction. The cell gap ( $\simeq$ 5 to 8 $\upmu$m) was maintained by spherical silica spacers mixed with UV curable optical adhesive (NOA-81). An inverted polarising optical microscope (Nikon eclipse Ti-U) with a 60X water immersion objective and a colour camera
(Nikon DS-Ri2) is used for texture observation and video recording. A first order waveplate, the so-called $\lambda$-plate (530 nm) is used for constructing director field ($\bf{n}$) surrounding the particles. A laser tweezer operating at 1064 nm (Tweez 250si) is built on the inverted microscope. An acousto-optic deflector (AOD) is used for optical trap and hence the particle manipulation. A charge-coupled device (iDs-UI) is used for the video recording of the particle trajectories. A particle tracking program is used off-line to track the centres of the particles with an accuracy of 20nm.

\section{Results and discussion}
\begin{figure}[ht]
\centering
\includegraphics[scale=0.32]{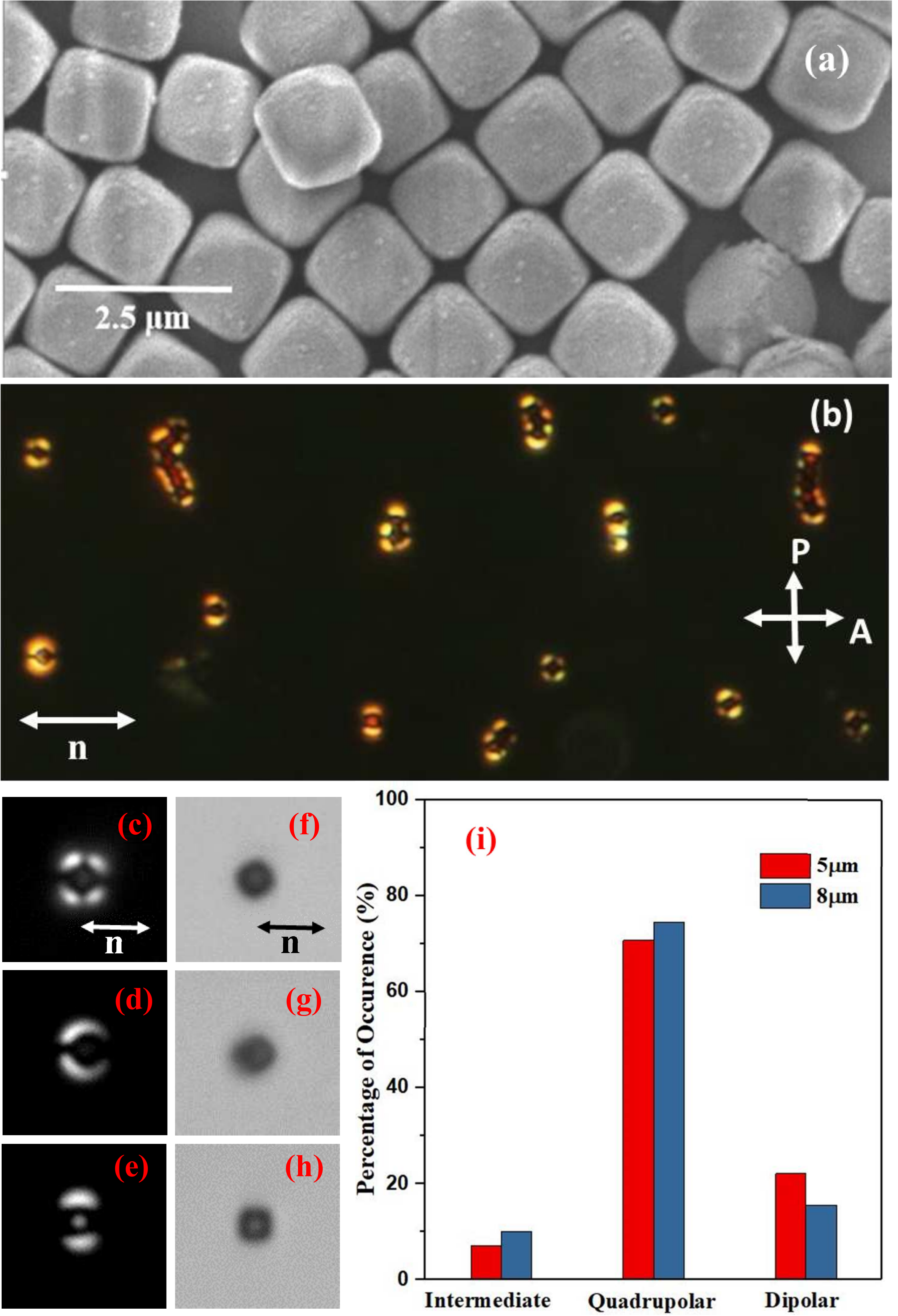}
\caption{(a) Scanning electron microscope (SEM) micrograph of the silica coated hematite microcubes. (b) Polarising optical microscope (POM) micrograph of dispersed microcubes in the nematic liquid crystal 5CB. POM micrographs of (c) quadrupolar (d) dipolar and (e) intermediate type elastic distortions. (f,g,h) Micrographs of the corresponding microcubes without polariser and analyser. (i) Percentage of occurrence of three types of elastic distortions. About 200 particles are studied in two different cells with thickness $5{\upmu}$m (red) and $8{\upmu}$m (blue).  Double headed arrows above $\bf{n}$ denote the director, which is parallel to the rubbing direction. } 
\label{fig:figure2}
\end{figure}
The microcubes are nearly mono-dispersed with slightly round edges as seen in the scanning electron microscope (SEM) image presented in Fig. \ref{fig:figure2}(a). The shape of our microcubes closely resembles the simulated microcubes with sharpness factor $m=3$ as shown in Fig. \ref{fig:figure1}. The average length of the face diagonal is about 1.8 $\upmu$m. Figure \ref{fig:figure2}(b) shows a polarising optical microscope (POM) micrograph with a few dispersed microcubes in 5CB liquid crystal. It is apparent that the microcubes have different orientations with respect to the rubbing direction, which can not be determined conspicuously due to the small size.  Nevertheless, careful observation at close-up reveals that there are mainly three types of orientations with distinct elastic distortions in which the first and second types can be identified as elastic quadrupoles (Fig. \ref{fig:figure2}(c)) and dipoles (Fig. \ref{fig:figure2}(d)); the third type (Fig. \ref{fig:figure2}(e)) exhibits an intermediate structure that does not resemble either of them. For a quadrupolar microcube, the face diagonal (Fig. \ref{fig:figure2}(f)), whereas the body diagonal of a dipolar microcube (Fig. \ref{fig:figure2}(g)) is parallel to the rubbing direction. In the intermediate structure, two opposite faces of the microcube are perpendicular to the rubbing direction (Fig. \ref{fig:figure2}(h)). Microscopic investigation on  a large number of microcubes shows that the majority ($\sim$80\%) of the microcubes are quadrupolar type and about $\sim$15\% of them are dipolar type (Fig. \ref{fig:figure2}(i)). There are only $\sim$5\% microcubes  exhibiting intermediate type director structure and the overall result is almost independent of the cell thickness.

\begin{figure}
\centering
\includegraphics[scale=0.23]{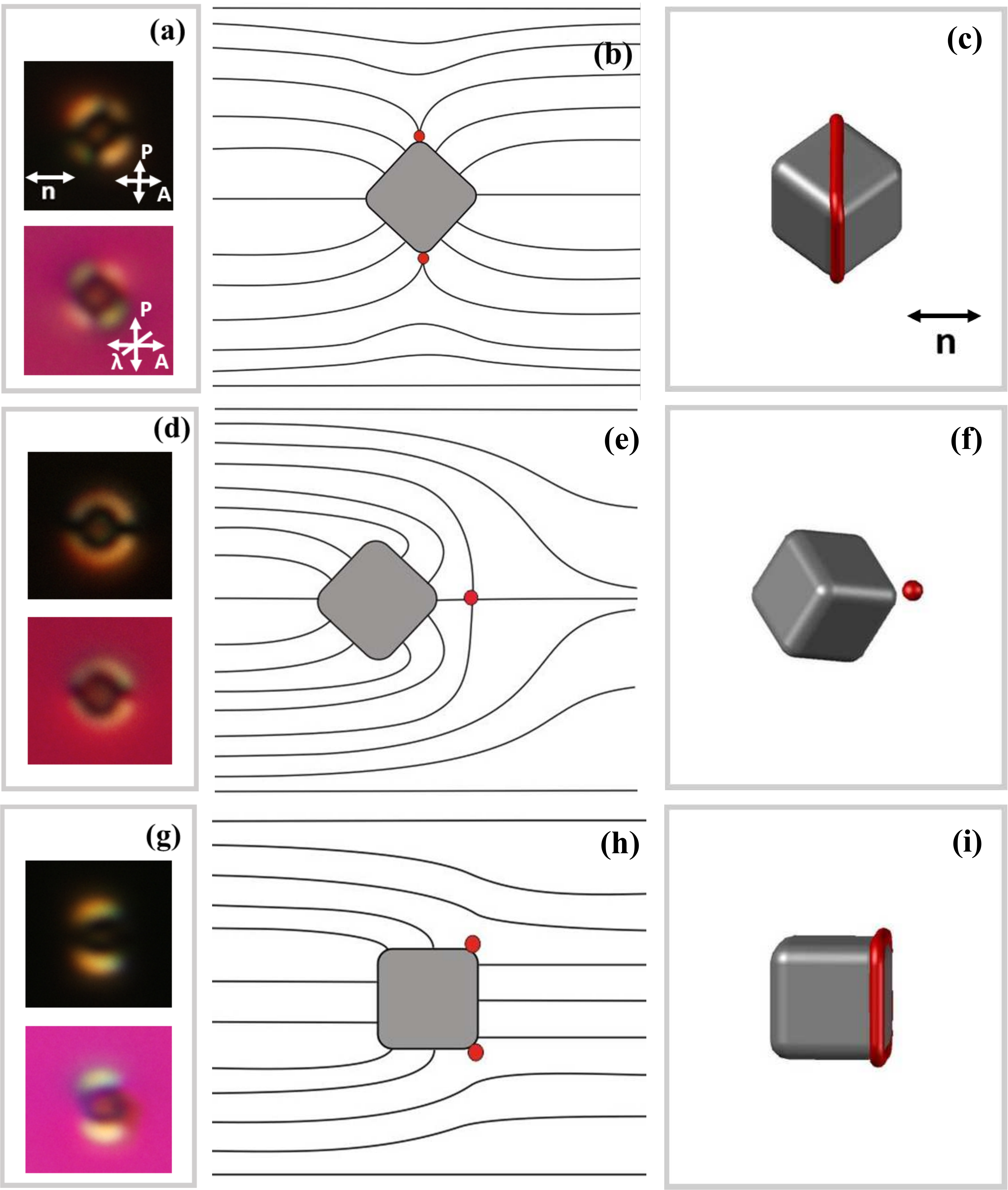}
\caption{(a,d,g) Polarising optical microscope (POM) and the corresponding $\lambda$-plate micrographs of three types of colloids.  (b,e,h) Constructed director profiles of each coloid. (c,f,i) Three dimensional orientation of the microcubes with induced defects. Defect rings (thick red lines) are pinned on the surface of the microcubes.} 
\label{fig:figure3}
\end{figure}

Qualitative information about the anchoring of the LC molecules and the resulting defect is obtained by inserting a full waveplate ($\lambda$-plate) at an angle of $45^{\circ}$ with respect to the crossed polarizers. The POM micrographs and the corresponding images taken with $\lambda$-plate are shown in Fig. \ref{fig:figure3}(a,d,g). The $\lambda$-plate introduces a phase shift of exactly one wavelength (530 nm) between the ordinary and extra-ordinary wavefronts as a result of which mainly three colours are observed. The magenta colour corresponds to either vertical or horizontal orientation of the director, whereas bluish and yellowish colours correspond to clockwise and anticlockwise rotation of the director, respectively\textsuperscript{\cite{utk,igor}} as shown in Fig. \ref{fig:figure3}(b,e,h). Three-dimensional schematic diagrams with  induced defects of the microcubes are also presented in Fig. \ref{fig:figure3}(c,f,i). Figure \ref{fig:figure3}(c) shows that two opposite face diagonals of the microcube are parallel to $\bf{n}$ and the disclination ring (Saturn ring)  encompasses the four vertices of a body diagonal plane. In Fig. \ref{fig:figure3}(f), one of the body diagonals is parallel to $\bf{n}$ and the microcube is accompanied by a point defect, closely resembling its spherical counterpart. The texture shown in Fig. \ref{fig:figure3}(g) evokes an interesting situation where the surrounding director profile as shown in Fig. \ref{fig:figure3}(h) is roughly in agreement with the dipole, except the point defect. This suggests that the intermediate texture corresponds to a defect structure, where the defect ring wraps the four edges of a face, which is perpendicular to the director $\bf{n}$  as shown in Fig. \ref{fig:figure3}(i). In this case, the face normals of two opposite faces are parallel to $\bf{n}$. 

\begin{figure}[ht]
\centering
\includegraphics[scale=0.25]{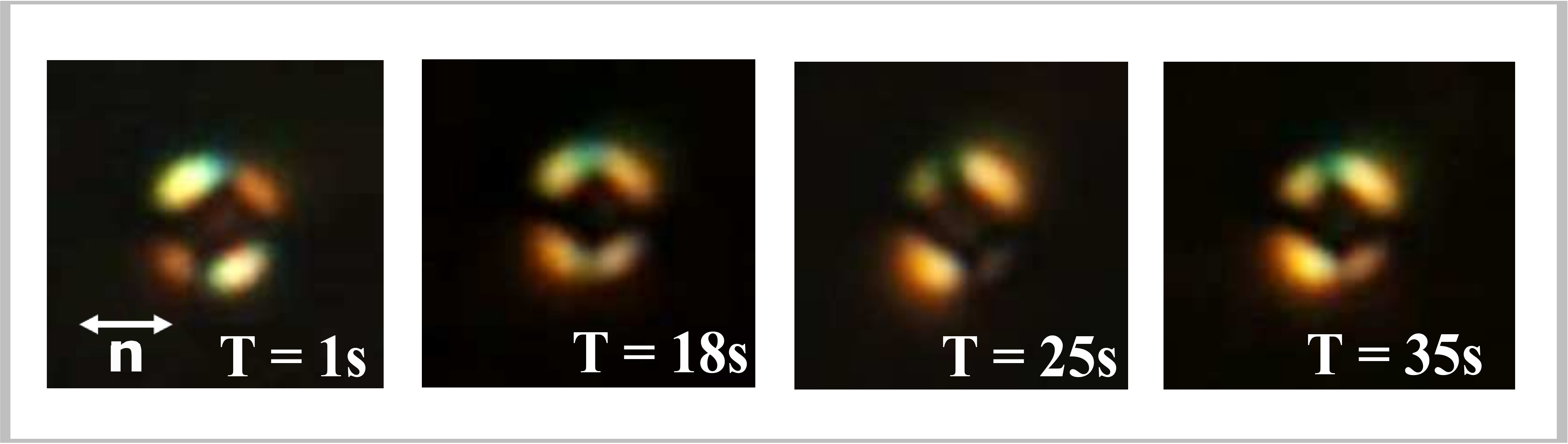}
\caption{Snapshots of rotational Brownian motion of a microcube at different time interval. (See Movies S1 and S2). } 
\label{fig:figure4}
\end{figure}

There are a very few computer simulation studies on the defects and pair interaction of cubic colloids using Landau-de Gennes (LdG) $Q$-tensor theory. Hung \textit{et al.} showed that cubic nano-colloids mostly stabilise quadrupolar structure with distorted disclination ring, encircling nanocubes, where the ring bends along six of the twelve edges of the cube\textsuperscript{\cite{hung}}.
Our experiments show that microcubes predominantly ($\sim$80\%) stabilise quadrupolar structure, which is consistent with the predictions of the simulation (Fig.\ref{fig:figure2}(i)). 
 Beller \textit{et al.} predicted the polymorphism of the disclination ring, showing the rewiring of the defect ring to a new set of edges when the cube is rotated about a vertical axis through its centre\textsuperscript{\cite{beller}}. They showed that all these defect configurations of the microcubes correspond to the degenerate states of the LdG free energy. However, experimentally it is difficult to rotate the cube as described in the simulation. Hence, we looked at the Brownian motion of a quadrupolar microcube. The microcube executes rotational diffusion as a result of which the quadrupolar texture evolves continuously with time (Movies S1 and S2). A few snapshots captured at different time intervals are shown in Fig. \ref{fig:figure4}. The temporal change of the texture is due to the rotation of the cube which could lead to the rewiring of the disclination ring as predicted\textsuperscript{\cite{beller}}. However, because of small size and the roundness of the edges, it is difficult to identify the different states with reconfigured disclination ring.

 \begin{figure}
\centering
\includegraphics[scale=0.2]{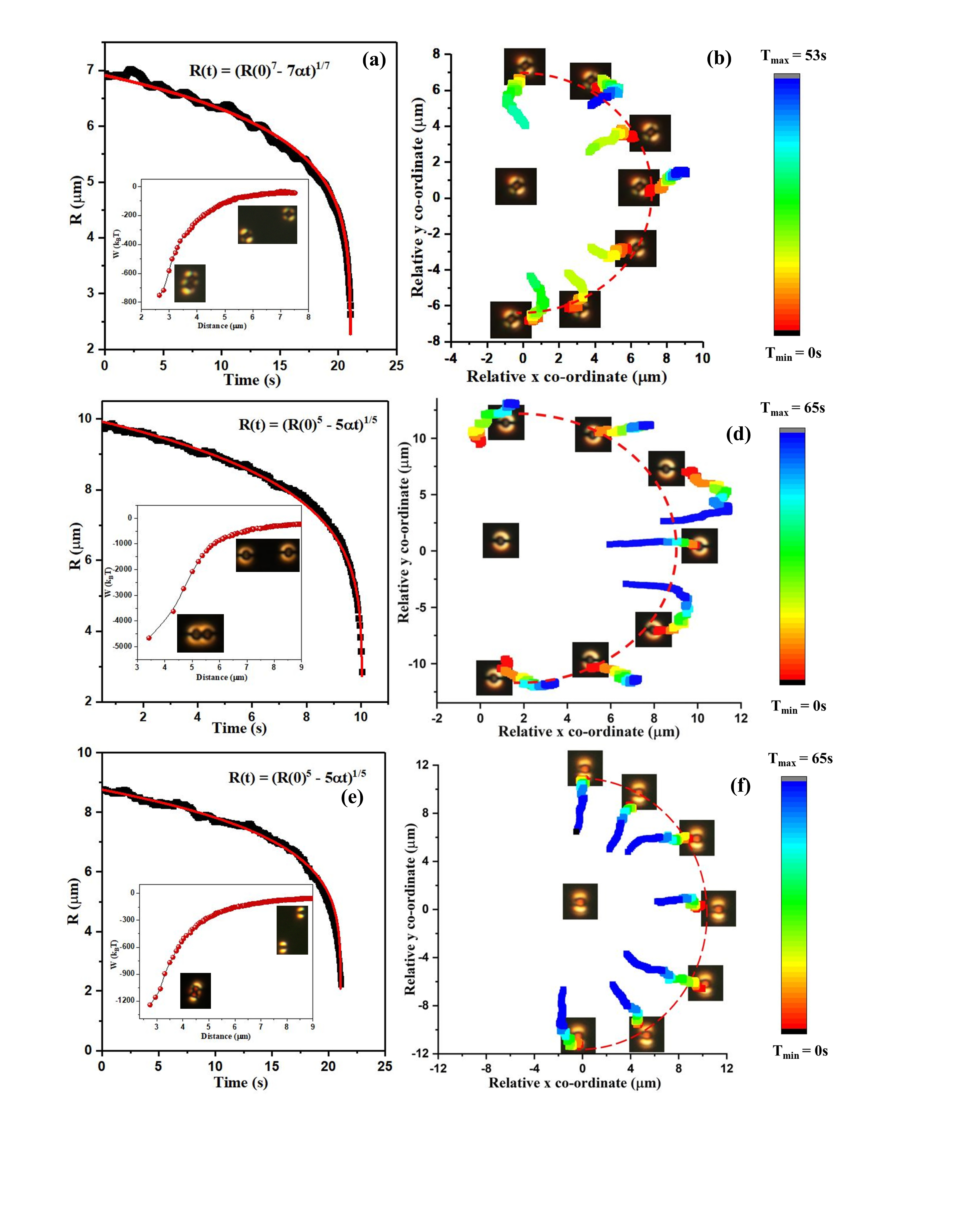}

\caption{Interparticle separation $R(t)$ with time of a pair of interacting (a) quadrupolar, (c) dipolar microcubes and (e) the microcubes exhibiting intermediate texture. Red lines are the least squares fits corresponding to equations $R(t)=\{R(0)^{7}-7{\alpha}t\}^{1/7}$,
for quadrupolar and $R(t)=\{R(0)^{5}-5{\alpha}t\}^{1/5}$, for dipolar interactions. Corresponding potential energies ($W(r)$) in units of k\textsubscript{B}T are shown in the inset. (b,d,f) Relative coordinates and resulting colour-coded trajectories as a function of time for respective colloids.  By ``relative coordinate'' we mean relative to the starting point of each trajectory. }
\label{fig:figure5}
\end{figure}

  \begin{figure}[ht]
\centering
\includegraphics[scale=0.25]{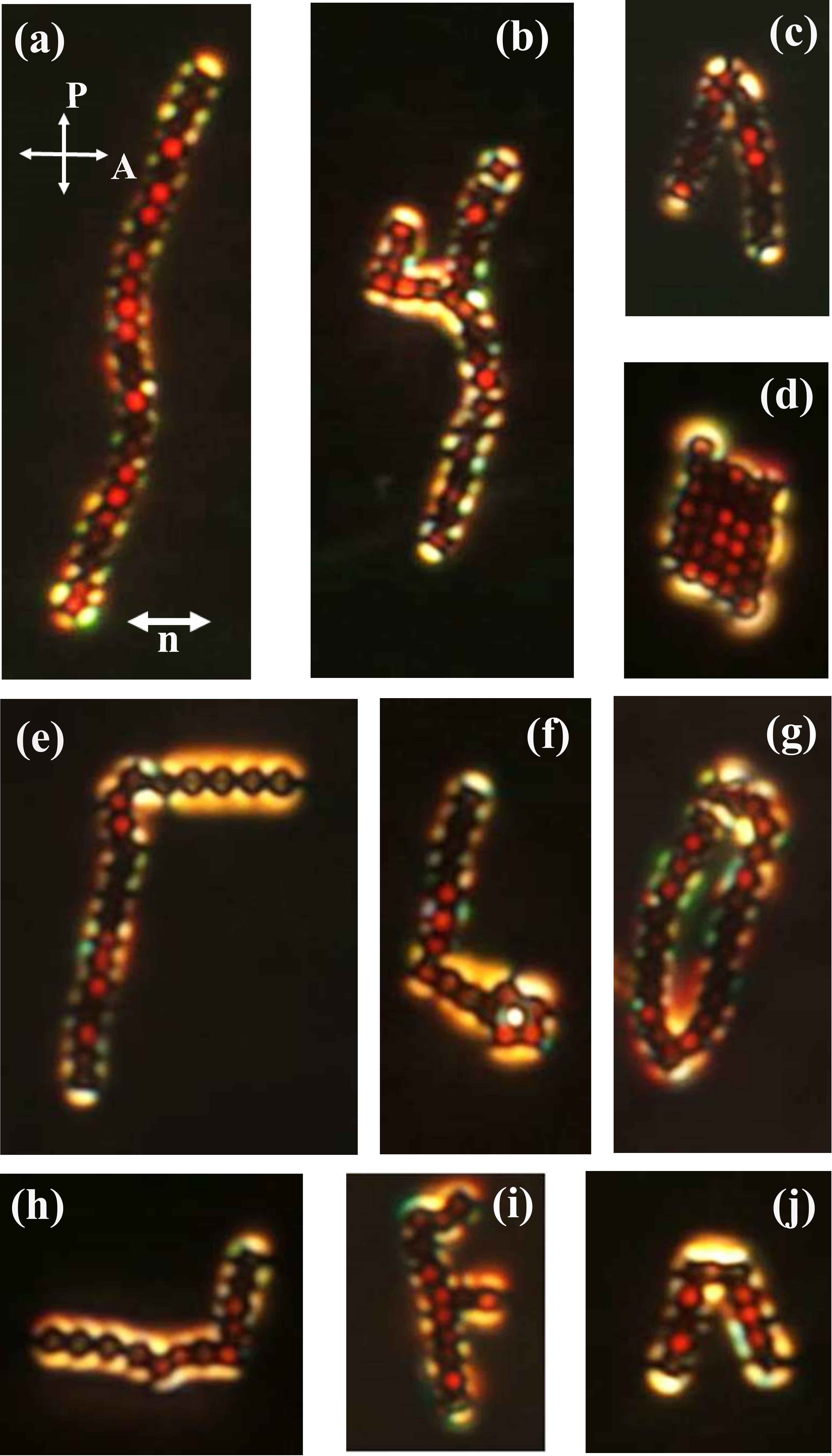}
\caption{ Laser tweezers assisted assembly of microcubes with (a,e,h) bent chains, (b,i) branches, (d,j) kinks, (d) 2D  crystal and (f,g) closed-loop structures. CCD images of the same are shown in the supplementary materials (Fig.S1)}
\label{fig:figure6}
\end{figure}

As a next step we study the elastic pair interaction of the microcubes using videomicroscopy technique. We trap two microcubes using the laser tweezers and release them from a few micrometers apart and record their motion. The motion of the microcubes in the NLC is overdamped due to the high viscosity, hence the net force experienced by the moving microcubes at any instant is zero. This means the elastic force $F_{el}$ is balanced by an opposing viscous drag force, $F_{drag}={-\zeta}dR/dt$, i.e., $\vec{F}_{el}+\vec{F}_{drag}=0$, where $\zeta$ is the drag coefficient, and $R(t)$ is the interparticle separation. The interaction potential $W(r)$ is obtained by integrating the drag force over the trajectory i.e., $W(r)=\int{\vec{F}_{drag}.\vec{dr}}$\textsuperscript{\cite{zumer}}. Using methods similar to those reported in ref.\textsuperscript{\cite{utk,rasi}}, we measure the drag coefficients $\zeta_{\parallel} \simeq 0.9\times 10^{-6}$ kg/s and $\zeta_{\perp} \simeq 1.7\times 10^{-6}$ kg/s, by measuring the anisotropic diffusion coefficients $D_{\parallel}$ and $D_{\perp}$ at room temperature ($T=298$ K). The subscripts refer in relation to the director ${\bf n}$. The drag coefficients weakly depend on the deformations hence, the average drag coefficient $\zeta = (\zeta_{\parallel} + \zeta_{\perp})/2 \approx 1.3\times 10^{-6} $ kg/s is used in calculating the interaction energy. The elastic force between two quadrupolar colloids is given by $F_{e}=-k/R^{6}$ and the corresponding inter-colloid separation is given by\textsuperscript{\cite{laponite,rasna}}
\begin{equation}
R(t)=\{R(0)^{7}-7\alpha t\}^{1/7},
\end{equation}
where $\alpha=K/\zeta$ and $R(0)$ is the initial separation at $t=0$ s.
Figures \ref{fig:figure5}(a) shows the variation of $R(t)$ of two quadrupolar microcubes and the nonlinear least squares fit to Eq.(1). The binding energy of two quadrupolar microcubes is $\sim$800 k\textsubscript{B}T (inset to Fig. \ref{fig:figure5}(a)). The angle dependence of interaction of the microcubes, presented in Fig. \ref{fig:figure5}(b), resembles to that of the spherical particles. For dipolar colloids the interaction is given by $F_{e}=-k/R^{4}$ and the time dependent inter-particle separation can be written as\textsuperscript{\cite{laponite,rasna}}
\begin{equation}
 R(t)=\{R(0)^{5}-5\alpha t\}^{1/5}.
 \end{equation}
 Figures \ref{fig:figure5}(c) shows the variation of $R(t)$ and the corresponding least squares fit to Eq.(2). The binding energy of two dipolar microcubes is $\sim$5000 k\textsubscript{B}T (inset to Fig. \ref{fig:figure5}(c)), which is much larger than that of quadrupolar microcubes. 
The interparticle separation $R(t)$ for two microcubes with intermediate structure (see Fig. \ref{fig:figure3}(g)) is shown in Fig. \ref{fig:figure5}(e). $R(t)$ fits well to Eq.(2) and the corresponding interaction energy is shown in the inset. The binding energy of two such microcubes is $\sim$1300 k\textsubscript{B}T (inset to Fig. \ref{fig:figure5}(e)), which is much lower than the dipolar microcubes. Moreover, the angle dependence of the interaction of microcubes with intermediate structure is markedly different than the dipolar microcubes, where the point defects of both the colloids are on the same side (Fig. \ref{fig:figure5}(d)). In particular, the interaction between two intermediate microcubes is attractive at all angles (Fig.\ref{fig:figure5}(f)), whereas for the dipolar microcubes, it is either attractive or repulsive, depending on the approaching angles with respect to the director (Fig.\ref{fig:figure5}(d)). This unusual behaviour of the microcubes with intermediate structure could be related to the relocation of the defect loop between the two orthogonal (with respect to $\bf{n}$) faces while interacting. Taking advantage of the anisotropic interactions and the cubic shape, we prepared several equilibrium assemblies of the microcubes with the help of the laser tweezers. Figure \ref{fig:figure6} shows a few nonlinear chains, branches, kinks, 2D crystals and closed-loop structures.  The assembled structures of the microcubes are diverse when compared to that of  
spherical colloids which primarily stabilise either linear or zigzag chains\textsuperscript{\cite{zumer}}. The segments which are parallel or perpendicular to the far field director are mostly composed of either dipolar and quadrupolar colloids, respectively. Microcubes with intermediate structure helps in forming branches, kinks and loops. These structures are highly stable against thermal fluctuations and respond to external magnetic fields, hence could be useful as building blocks in making advanced magnetic materials with desired architecture. 
 
  \begin{figure}[ht]
\centering
\includegraphics[scale=0.2]{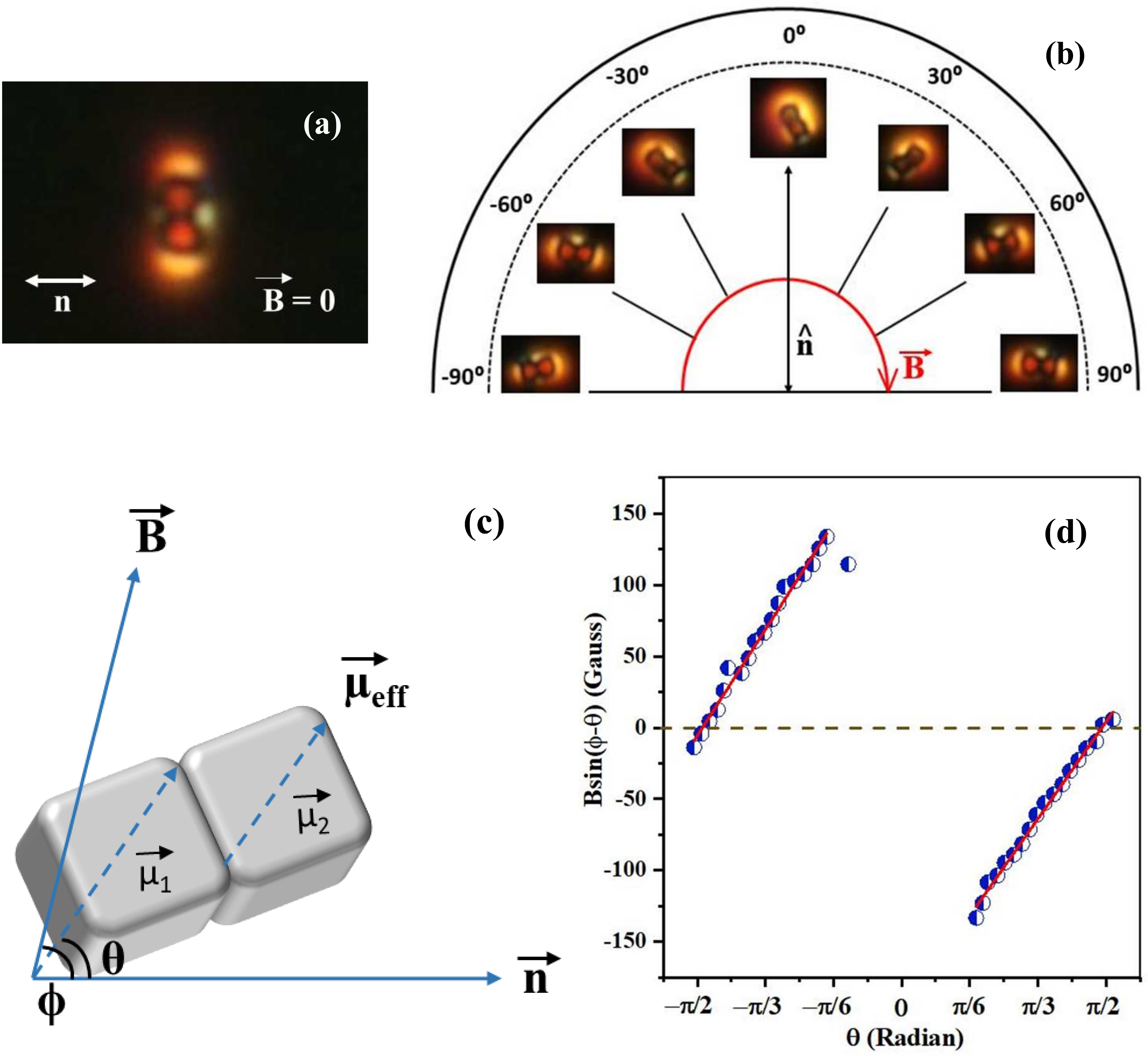}
\caption{ (a) POM micrograph of a pair of assembled quadrupolar cubes in the absence of magnetic field ($\vec{B}=0$). (b) The direction of $\vec{B}$ is  changed from $-90^{\circ}$ to $+90^{\circ}$ as shown by the red-arrow, keeping the field strength fixed at 300 Gauss. POM micrographs of the reoriented pair at equilibrium are also shown. The pair flips the orientation at $\theta=0$. (c) Individual magnetic moments $\vec{\mu}_1$ and $\vec{\mu}_2$ of the microcubes and the resultant moment $\vec{\mu}_{eff}$ of the pair. $\theta$ and $\phi$ are the angles subtended by $\vec{\mu}_{eff}$ and $\vec{B}$ with respect to $\bf{n}$. (d) Variation of $B\sin({\phi}-{\theta}$) with ${\theta}$. The red lines are the linear least squares fits to the equation: $B\sin({\phi}-{\theta})=(4{\pi}CK/\mu_{eff}) \theta$ with an average slope of 143 Gauss/rad.}
\label{fig:figure7}
\end{figure}

\begin{figure}[ht]
\centering
\includegraphics[scale=0.2]{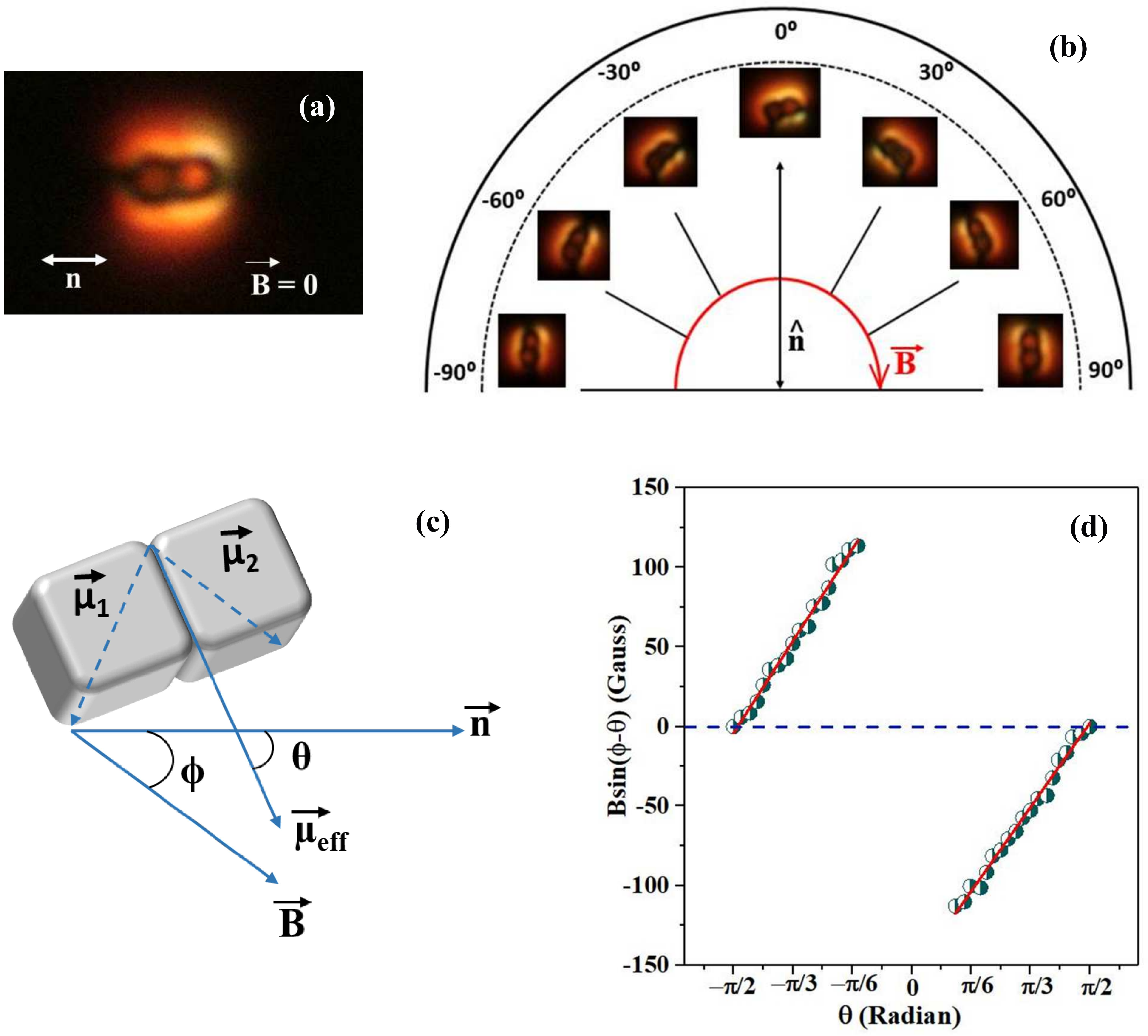}
\caption{(a) POM micrograph of a pair of assembled dipolar microcubes in the absence of magnetic field ($\vec{B}=0$). (b) $\vec{B}$ is  changed from $-90^{\circ}$ to $+90^{\circ}$ as shown by the red arrow, keeping the strength fixed at 300 Gauss. POM micrographs of the reoriented pair are also shown. The pair flips the orientation at $\theta=0$. (c) Schematic representation of the direction of individual magnetic moments $\vec{\mu}_1$ and $\vec{\mu}_2$ of the microcubes and the resultant moment $\vec{\mu}_{eff}$ of the pair. $\vec{\mu}_{eff}$ and $\vec{B}$ makes angle $\theta$ and $\phi$ with respect to $\bf{n}$. (d) Variation of $B\sin({\phi}-{\theta}$) with ${\theta}$. The red lines are the linear least squares fits to the equation: $B\sin({\phi}-{\theta})=(4{\pi}CK/\mu_{eff})\theta$ with an average slope of 106 Gauss/rad. }
\label{fig:figure8}
\end{figure}

The microcubes possess a permanent magnetic moment along a direction which is nearly parallel to the body diagonal\textsuperscript{\cite{rossi1}}. In water, Earht's magnetic field helps in forming linear chains, which are parallel to the field direction\textsuperscript{\cite{jm,rossi2}}. However, in NLC, the orientation of the microcubes is mainly governed by the surface anchoing and the ensuing energy of the elastic distortion and is unaffected by the Earth's magnetic field. To study the effect of applied magnetic field, we used two disc-shaped magnets and positioned them diametrically opposite on a circular track around the plane of the sample as shown in the supplementary materials (Fig.S2). This arrangement provides an in-plane magnetic field which can be rotated around the cell by simultaneously rotating the two magnets. The strength of the field at the sample position measured by a magnetometer is 300 Gauss. Although the individual microcubes respond to the rotating magnetic field, it is difficult to determine the angle of rotation conspicuously. Hence, we assembled dipolar and quadrupolar pairs of the microcubes using the laser tweezers and studied their magnetic response. Figure \ref{fig:figure7}(a) shows a pair of quadrupolar microcubes, oriented perpendicular to $\bf{n}$ in the absence of any magnetic field. When the magnetic field is rotated from $-90^{\circ}$ to $+90^{\circ}$ around the director, the body diagonal of the pair tends to follow the field direction as shown in Fig. \ref{fig:figure7}(b). This means the direction of the resultant magnetic moment ($\vec{\mu}_{eff}=\vec{\mu}_1+\vec{\mu}_2$) is approximately parallel to the body diagonal of the pair as shown in Fig.\ref{fig:figure7}(c). For a dipolar pair, the long body axis is parallel to the director in the absence of magnetic field as shown in Fig. \ref{fig:figure8}(a). When the magnetic field is rotated the short body axis of the pair tends follow the field direction (Fig. \ref{fig:figure8}(b)). This suggests that the direction of the resultant moment $\vec{\mu}_{eff}$ is along the short body axis of the pair as shown in Fig. \ref{fig:figure8}(c).

The rotation of the microcube-pair increases the elastic energy, which opposes the effect of the magnetic torque $\vec{\mu}_{eff}\times\vec{B}$. In both  cases, $\vec{\mu}_{eff}$ and the magnetic field $\vec{B}$ makes angle $\theta$ and $\phi$ with respect to the director $\bf{n}$. At an equilibrium, the magnetic torque $\mu_{eff}B \sin(\phi-\theta)$ is balanced by the gradient of the elastic energy $-\partial U/\partial\theta$ and the torque balance equation can be written as\textsuperscript{\cite{Brochard,Lapointe2}}: 
\begin{equation}
-\partial U/\partial\theta+\mu_{eff}B \sin(\phi-\theta)=0.
\end{equation}
The enhancement in the elastic energy due to the rotation by an angle $\theta$ is given by $U=2{\pi}CK{\theta}^{2}$, where $K$ is the effective Frank elastic constant and $C=L/[2\text{log}(L/d)]$, where $L$ and $d$ are the length and diameter of the assembled pair. The magnetic moments of both the pairs are determined from the slope of the equation $B\sin({\phi}-{\theta})=(4{\pi}CK/\mu_{eff})\theta$. Figures \ref{fig:figure7}(d) and \ref{fig:figure8}(d) shows that $B\sin({\phi}-{\theta})$ is linearly proportional to $\theta$ with slopes 143 and 106 Gauss/rad for quadrupolar and dipolar pairs, respectively. Taking $C= 4.0\times 10^{-6}$ m and $ K=5\times 10^{-12}$ N, the estimated magnetic moments for quadrupolar and dipolar pairs are given by  ${\mu}_{eff}\simeq 18\times 10^{-15}$  Am\textsuperscript{2} and $24\times 10^{-15}$ Am\textsuperscript{2} respectively. Assuming that both the magnitudes are identical, the magnetic moment of a single microcube in quadrupolar pair is given by $\mu_{1}=\mu_2={\mu}_{eff}/2\simeq 9 {\times} 10^{-15}$ Am\textsuperscript{2}. For the dipolar pair, the magnetic moment along the body diagonal makes an angle of 35$^\circ$ with respect to the face diagonal and hence the magnetic moment is given by $\mu_{1}=\mu_2= {\mu}_{eff}/2\cos(35^{\circ})\ \simeq 14{\times} 10^{-15}$ Am\textsuperscript{2}. The average magnetic moment obtained from two differently assembled pairs of microcubes are given by $\mu\approx 11.5 {\times} 10^{-15}$ Am\textsuperscript{2}. This result is very close to the theoretically calculated and experimentally measured value of the bulk sample composed of similar microcubes by an alternating current gradient magnetometer\textsuperscript{\cite{rossi1}}. However, our method is very simple and provides measurement at microscopic level than the conventional method used for bulk sample composed of a very large number of microcubes.

\section{Conclusion} 
     Microcubes with homeotropic surface anchoring exhibit mainly three different orientations in a nematic liquid crystal. For majority of the microcubes, the face diagonals of a pair of opposite faces are parallel to the rubbing direction and exhibit quadrupolar interaction. Dipolar structure with a point defect is exhibited by a small number of microcubes for which one of the the body diagonals is nearly parallel to the rubbing direction. In a very small fraction of microcubes, two opposite faces are oriented perpendicular to the rubbing direction in which the disclination ring wraps around the four edges of either of the faces. The angle dependence of the elastic interactions are markedly different for all three distinct orientations of the microcubes. Our experiments provide an evidence of defect polymorphism of the microcubes, which was predicted earlier by the computer simulation. With the help of the laser tweezers, we assembled microcubes into various structures, such as highly bent chains, branches, kinks, closed loops, which are not achievable in spherical colloids. The combined effects of particle-shape, anisotropic interaction and magnetic properties of the microcubes in liquid crystals are effectively used to design diverse building blocks which could be useful for assembling complex, programmable and magnetically responsive colloidal structures. Our investigations have focused on cubic particles; however, the plethora of faceted colloids accessible promises a wide range of as yet unexplored phenomena and their applications.     \\\\\\\\\
     
  \newpage   
{\bf Supplementary Materials}\newline
Supplementary materials are available online or from the author.\\

{\bf Acknowledgments}\newline
SD acknowledges the support from DST (DST/SJF/PSA-02/2014-2015) and DST-PURSE. DVS acknowledges DST for INSPIRE fellowship. RKP acknowledges DST for INSPIRE Faculty award grant (DST/INSPIRE/04/2016/002370).\\
  
{\bf Conflict of Interest}
\newline
The authors declare no conflict of interest.\\

{\bf Keywords}
\newline
Cubic hematite colloids, Topological defects, Self-assembly, Magnetic moments\\

\end{document}